\documentclass[sigconf]{acmart}
\AtBeginDocument{%
  \providecommand\BibTeX{{%
    \normalfont B\kern-0.5em{\scshape i\kern-0.25em b}\kern-0.8em\TeX}}}

\usepackage{booktabs} % For formal tables
\usepackage{graphicx}
\usepackage{balance}
\usepackage{pifont} % for \ding
\usepackage{longtable}
\usepackage{enumitem} % for myitemize and myenumerate
\usepackage[algoruled, linesnumbered, vlined]{algorithm2e}
\usepackage{algorithmic}
\usepackage{setspace}
\usepackage{multirow}
\usepackage{url}
\usepackage[T1]{fontenc}

\usepackage{sparklines}
\usepackage{xspace}
\usepackage{fancyvrb}
\usepackage{xcolor}
\usepackage{soul}
\usepackage{marginnote}
\newcommand{\ie}{i.e.,\ }
\newcommand{\eg}{e.g.,\ }

\newcommand{\theproject}{Data Station}

\newcommand{\tinyskip}{\vspace{3pt}}
\newcommand{\mypar}[1]{\tinyskip\noindent\textbf{#1.}\xspace}

\newcommand{\F}{\mbox{Fig.\hspace{0.25em}}}

\newenvironment{myitemize}{%
\begin{itemize}[leftmargin=1em, itemsep=.1em, parsep=.1em, topsep=.1em,
    partopsep=.1em]}
{\end{itemize}}

\newenvironment{myenumerate}{%
\begin{enumerate}[leftmargin=1em, itemsep=.1em, parsep=.1em, topsep=.1em,
    partopsep=.1em]}
{\end{enumerate}}

\newenvironment{structure*}{\color{blue}\begin{myenumerate}}{\end{myenumerate}}

\sethlcolor{yellow}
\newif\iffinal
\iffinal
  \newcommand\ian[1]{}
  \newcommand\raul[1]{}
  \newcommand\ben[1]{}
  \newcommand\kyle[1]{}
\else
  \newcommand\ian[1]{\textcolor{blue}{Ian: #1}} 
  \newcommand\raul[1]{\textcolor{purple}{Raul: #1}} 
  \newcommand\ben[1]{\textcolor{green}{Ben: #1}}
  \newcommand\kyle[1]{\textcolor{red}{Kyle: #1}}
\fi

\setcopyright{acmcopyright}
\copyrightyear{2018}
\acmYear{2018}
\acmDOI{10.1145/1122445.1122456}

\acmConference[CIDR'21]{CIDR'21 
Conference on Innovative Data Research}{January 03--05, 2021}{Asilomar, CA}
\acmBooktitle{CIDR'21}
\acmPrice{15.00}
\acmISBN{978-1-4503-XXXX-X/18/06}

\begin{document}

%\title{The Data Station: Next Generation Data Lakes}
%\title{The Data Station: Next Generation Data Architectures}
%\title{The Data Station: Centralizing Data and Computation}
%\title{The Data Station: Data, Computation, and Market Forces}
%\title{The Data Station: Data, Compute, and Market Forces, Together}
\title{The Data Station: Combining Data, Compute, and Market Forces}

% vldb thing
%\numberofauthors{1}

\newcommand{\uc}{*}
\newcommand{\ucb}{\bullet}
\newcommand{\anl}{\dagger}

% SIGCONF version
\author{Raul Castro Fernandez$^\uc$, Kyle Chard$^{\uc\anl}$, Ben Blaiszik$^{\uc\anl}$, Sanjay Krishnan$^\uc$, Aaron
Elmore$^\uc$, Ziad Obermeyer$^\ucb$, Josh Risley$^\uc$, Sendhil Mullainathan$^\uc$, Michael Franklin$^\uc$,
Ian Foster$^{\uc\anl}$}
\email{[raulcf,chard,blaiszik,skr,aelmore,mjfranklin,foster,josh.risley,mullainathan]@uchicago.edu,
zobermeyer@berkeley.edu}
%\author{G.K.M. Tobin}
\affiliation{%
  \institution{$^\uc$The University of Chicago, $^\ucb$The University of California at
Berkeley, $^\anl$Argonne National Laboratory}
%\email{webmaster@marysville-ohio.com}
%\affiliation{%
%  \institution{Institute for Clarity in Documentation}
%  \streetaddress{P.O. Box 1212}
%  \city{Dublin}
%  \state{Ohio}
%  \postcode{43017-6221}
}

\renewcommand{\shortauthors}{Raul Castro Fernandez, Kyle Chard, and others}

\begin{abstract}

This paper introduces Data Stations, a new data architecture that we are
designing to tackle some of the most challenging data problems that 
we face today: access to sensitive data; data discovery and integration;
and governance and compliance. Data Stations depart from modern data lakes in
that both data and derived data products, such as machine learning models, are
sealed and cannot be directly seen, accessed, or downloaded by anyone. Data
Stations do not deliver data to users; instead, users bring questions to data.
This inversion of the usual relationship between data and compute mitigates many
of the security risks that are otherwise associated with sharing and working
with sensitive data.

Data Stations are designed following the principle that many data problems
require human involvement, and that incentives are the key to obtaining such
involvement. To that end, Data Stations implement market designs to create,
manage, and coordinate the use of incentives. We explain the motivation for this
new kind of platform and its design.

\end{abstract}

\maketitle

\section{Introduction}

Whenever data and models are shared, transformation ensues. Breaking down data silos
unleashes value that makes companies more competitive. Pooling knowledge, such as when
hospitals form coalitions, accelerates discovery. Entire disciplines
change when researchers share benchmarks and models~\cite{medicalimaging,nlu}.
%Whenever data and models are shared, transformation ensues.  For example,
%companies are transformed when data silos are broken down; scientific discovery
%is accelerated when hospital coalitions pool knowledge; and entire research
%domains change when researchers share benchmarks and
%models~\cite{medicalimaging, nlu}. 
%%Companies change by
%%multiplying data's value when they break down data silos. Hospital coalitions 
%%accelerate scientific discovery by pooling knowledge. Entire
%%disciplines change when researchers share benchmarks and
%%models~\cite{medicalimaging, nlu}. 
However, three barriers prevent effective sharing:
\emph{easy access to sensitive data}, \emph{data discovery and integration}, and
\emph{data governance and compliance} are all challenges with both technical and
human components. 

Much prior work has tackled each barrier individually. However, 
individual solutions are often in conflict. 
For example, it is harder to discover relevant datasets when access is restricted,
and to govern data when underlying datasets are not well integrated. 
We need a comprehensive solution that addresses all three barriers together.

\mypar{Discovery and Integration} Data lakes~\cite{lake1, commons} ease data
access by collecting unrestricted datasets in a central repository where
they may be accessed and downloaded by analysts. 
However, large volumes of data mean analysts
spend more time in finding (discovery) and combining (integration) datasets than
in their analysis~\cite{eighty}. 

\mypar{Access to Sensitive Data} Organizations are wary of sharing data because
they fear information leakage~\cite{foster2018research}. Simple anonymization
techniques do not suffice~\cite{latanya, netflix}.  
These disincentives block data sharing and stymie innovation.

\mypar{Data Governance and Compliance} Analysts routinely download datasets from
databases to produce machine learning (ML) models, reports, and other
derived data products. The consequence is a governance nightmare for those who
want to control access to sensitive information, need to comply with regulations
such as GDPR~\cite{gdpr} and CCPA~\cite{ccpa}, or want to ensure ethical use of
data. 

To tackle these challenges, a radically new data architecture is needed
to address both the \emph{technical} and the \emph{human} problem. 
Such an architecture must change how people access, and use data.

\mypar{Enter the Data Station} In the Data Station architecture, both data and
derived data products---such as ML models, query results, and
reports---are sealed and cannot be directly seen, accessed, or downloaded by
anyone. The key idea is that instead of delivering data to users, users bring
questions to data. For example, instead of downloading a dataset to
train a ML model, a user may tell the Data Station what model they need
and the Station identifies a suitable data + model combination, 
trains the model on the data, and makes the trained model available for inference.
This inversion of compute and data mitigates many security risks of
sharing sensitive data. 

Centralizing data and computation permits fine-grained yet scalable data access:
users see results of their tasks only after they have been given permission. In
this model, data lifecycles and provenance are known, which permits straightforward
implementation of data governance policies. 
For example, it is possible to prohibit the use
of non-interpretable ML models; to control the attributes included in training data
to avoid propagating biased and unfair models; and to limit the data used for
deriving data products to avoid leaking sensitive data. In general, it is
possible to control \emph{what} and \emph{how} derived data products are
produced and used. 

Centralizing data and computation has another benefit: the Station sees all
datasets, all models, and all compute requests. This information lays the foundation
for the design of data markets~\cite{dmms}. Data markets incentivize humans to
share data and concentrate their effort where it matters most: assisting with
discovery and integration tasks. Market forces can be used to recruit humans to clean datasets,
to indicate how to join datasets, or to annotate datasets with tags and other
documentation.

%With the \theproject{} architecture, organizations will be able to expand their
%chain of trust to include the Station, knowing that nobody can access those data
%in ways that they have not explicitly permitted. 

\mypar{Data Stations differ from Data Federation Architectures}  Data
federation architectures allow access to disparate sources through common
schemas (global-as-view~\cite{discoveryambiguity} or others~\cite{garlic1,
garlic2}). Each organization controls its own data locally, and must arbitrate
query execution and release of results. Modern federated systems
use statistical database privacy techniques to control the release of
results~\cite{jenniesmc, federated1, federated2}.  Data Stations explore a
different point of the design space. By inverting data and compute, they escape
the need for a common schema, avoiding the agreement problem of data integration 
and opening up possibilities for computation beyond
relational queries. By sealing all data and derived data products, they maintain
the same level of security, but facilitate the enforcement of fine-grain access
and governance policies, and enable the implementation of market mechanisms.  
Data Stations mitigate risks, but they also introduce four new challenges:

%\begin{myitemize} 
$\bullet$ \textbf{C1. Data-Compute Inversion.} Data Stations users must submit computational requests 
(\eg queries, model training/inference, data preparation) without seeing
the data that their requests will engage. Methods are needed to allow users to determine if a dataset
is suitable for their needs or if they can trust derived data products in the
absence of crucial metadata (e.g., creator and provenance).  

$\bullet$ \textbf{C2. Data Discovery and Integration.} Upon receiving a computational request, the
Station must determine what datasets are needed to perform the task (discovery)
and how to prepare and combine those datasets to enable the computation (integration).

$\bullet$ \textbf{C3. Unbounded computation.} Unlike traditional data processing
platforms, in which computation is specified over concrete datasets, in the
\theproject{} model one does not know the appropriate dataset a priori. Thus,
the \theproject{} architecture introduces new resource management problems that
modern schedulers are not designed to solve.

$\bullet$ \textbf{C4. Data governance and access.} Centralization creates
opportunities for precise data governance; exploiting those opportunities requires efficient and secure fine-grained data access
management.
Interfaces are needed for declaring access and governance protocols, and an
engine is needed to control and audit enforcement.
%\end{myitemize}

To address these challenges Stations introduce the concept of a
\emph{data-unaware task capsule} to help users declare a computation without
seeing its data; a \emph{discovery and integration platform} that finds
and combines datasets to satisfy a capsule's request for computation; a new
scheduler and \emph{compute substrate} to map computations to compute resources; and support
for the definition and enforcement of governance and access policies---all powered
by a \emph{data asset catalog}. Solving the four challenges automatically
remains largely impossible in the general case without human guidance. To engage
humans where they are most needed, we design and manage incentives by using
market forces, making the Data Station an implementation of a data market
management system~\cite{dmms}.

The rest of this paper is as follows. In Section~\ref{sec:usecases} we
review use cases within and across organizations. We describe the Data Station in
Section~\ref{sec:architecture} and its
incentive mechanisms in Section~\ref{sec:market}. Finally, we review related
work and our contributions. %~\ref{sec:relatedwork}.

\section{Next Platform Requirements}
\label{sec:usecases}

We illustrate data problems \emph{within} and \emph{across}
organizations and define the requirements that we use to motivate the
architecture of Data Stations in the next section.

\subsection{Data Problems within Organizations}

In 2019, 20\% of managers from top companies claimed that the planned to
deploy ML technology, and 27\% that they had already implemented such
technologies. These numbers dropped to 4\% and 18\% in 2020~\cite{pwcsurvey}.
There are two main reasons for this trend. First, finding ML experts to make
use of available data is hard. Second, \emph{data} in these companies
are spread across heterogeneous repositories, such as databases, warehouses,
spreadsheets, and lakes, and are managed by different teams, departments, and
divisions. Managers, analysts, and stewards struggle to identify and prepare the
data required by downstream ML tasks when those data are stored in silos. \textbf{REQ1.
Management of data lifecyles} is necessary to discover relevant data.
\textbf{REQ2. Easy integration of data assets} 
is needed by data consumers to save time.

Accessing data requires engaging with IT department administrators who
enforce access controls and are in charge of ensuring that approved 
requests comply with regulations. 
Organizations may also want to implement other policies: for example, preventing
analysts from using data columns that represent a protected class as input to ML
models, or from using ML models that are not interpretable (under some
well-specified definition of interpretability) and that would produce data whose origin
cannot be easily explained. These governance needs cannot easily be met without a
platform that \textbf{REQ3. Allows users to declare and enforce governance
policies}.

\subsection{Sharing Data Across Organizations}

While data sharing among organizations who own complementary data has the promise of
producing combinatorial value, it is often prevented by the fear of leaking sensitive
or confidential information. We illustrate this opportunity with two 
use cases.

\mypar{Data-Driven Physiology} Consider the task of linking ECG waveform
patterns to sudden cardiac death, which kills 300,000 Americans every year.
Health researchers have identified a number of clinical risk factors (heart
failure, family history, etc.)---yet the vast majority of deaths occur in those
without any of these conditions. Machine learning could be used to identify waveform
signatures indicating elevated risk that could then be used to target preventive interventions. Similar exercises could yield insights into many other conditions.
The few studies
that have used small, proprietary datasets have given reason for optimism.
Unfortunately, health data is stuck in organizations that are wary of sharing it
for fear of leaking sensitive information. Researchers are forced to build
personal relationships with data providers, agree on formats and integration
strategy, and negotiate one-off data-sharing agreements, hence slowing
down innovation.

\mypar{Accelerating Materials Design and Discovery} The global advanced
materials market is forecast to reach \$2T by 2024~\cite{materialsmarket}. An
important component for innovation is data.  Materials science databases contain
large volumes of data that introduce challenges related to discovery,
integration, and sharing of potentially company-sensitive information. Examples
of such datasets may include curated materials properties extracted from
literature, corpora of experimental or simulated materials properties, and
results from multi-fidelity simulations from the atomic to macroscopic (e.g.,
density functional theory, molecular dynamics, finite element method). For
example, the Materials Genome Initiative~\cite{white2012materials,
blaiszik2019data} has fueled innovations on microelectronics, aerospace,
automotive, defense, energy, and health sectors. These data may be difficult
for any one team or company to collect, and may require large expenditures of
effort in experiment, simulation and curation. 

%\smallskip

\textbf{REQ4. Pooling data across organizations securely} is crucial for
researchers in medicine, materials science, and others, to bootstrap their
data-driven discovery and modeling efforts, reduce experimental and
computational costs, and spur new innovations. But having a technical solution
to sharing data is not sufficient. Data participants must be incentivized to
share data in a way that eases its utility to others. Data Stations
\textbf{REQ5. Implement market mechanisms to manage incentives}, so they
concentrate resources and time where it matters most. 
%Next, we introduce Data
%Stations and explain how their design addresses \textbf{REQ1-5}.

\section{Centralizing Data and Models}
\label{sec:architecture}

We present the Data Station architecture in
Section~\ref{subsec:overview} and describe in Sections~\ref{subsec:1}--\ref{subsec:unbounded} its major
components. We conclude in Section~\ref{subsec:summary} 
with a summary of how the architecture addresses the
challenges (C1-C4) and requirements (REQ1-REQ5).

\subsection{The Data Station Architecture} 
\label{subsec:overview}

%\begin{figure*}[h] \centering
%\includegraphics[width=\textwidth]{img/architecture_white.pdf} 
%\caption{Center shows data Station architecture. Left hand side shows example of
%data delivery and associated access control.  Right hand side shows example of
%task capsule definition. Boxes shaded in red indicate ongoing work (see
%Section~\ref{sec:deliver}).
%\notera{we need a more detailed arch figure with
%interactions if possible}} 
%\label{fig:architecture} 
%\end{figure*}

We differentiate between \emph{data contributors}, who deliver data to the
\theproject{}, and \emph{data users}, who use these data to solve problems. We
talk about \emph{original data} or \emph{datasets} to refer to content that
contributors deliver to the platform, and \emph{derived data product} to refer
to datasets, models, visualizations, reports, or any other result obtained by
processing an original dataset. Data contributors use an interface to deliver
data securely to the Station, much as they interact with data lakes
today. Once data enters the Station, they are sealed and nobody can access them, or
any data product derived from those data, directly. 
We next explain how the Station is used from the perspectives of first a data user and then 
a data
contributor. 

\begin{figure}[t] 
\centering
\includegraphics[width=\columnwidth]{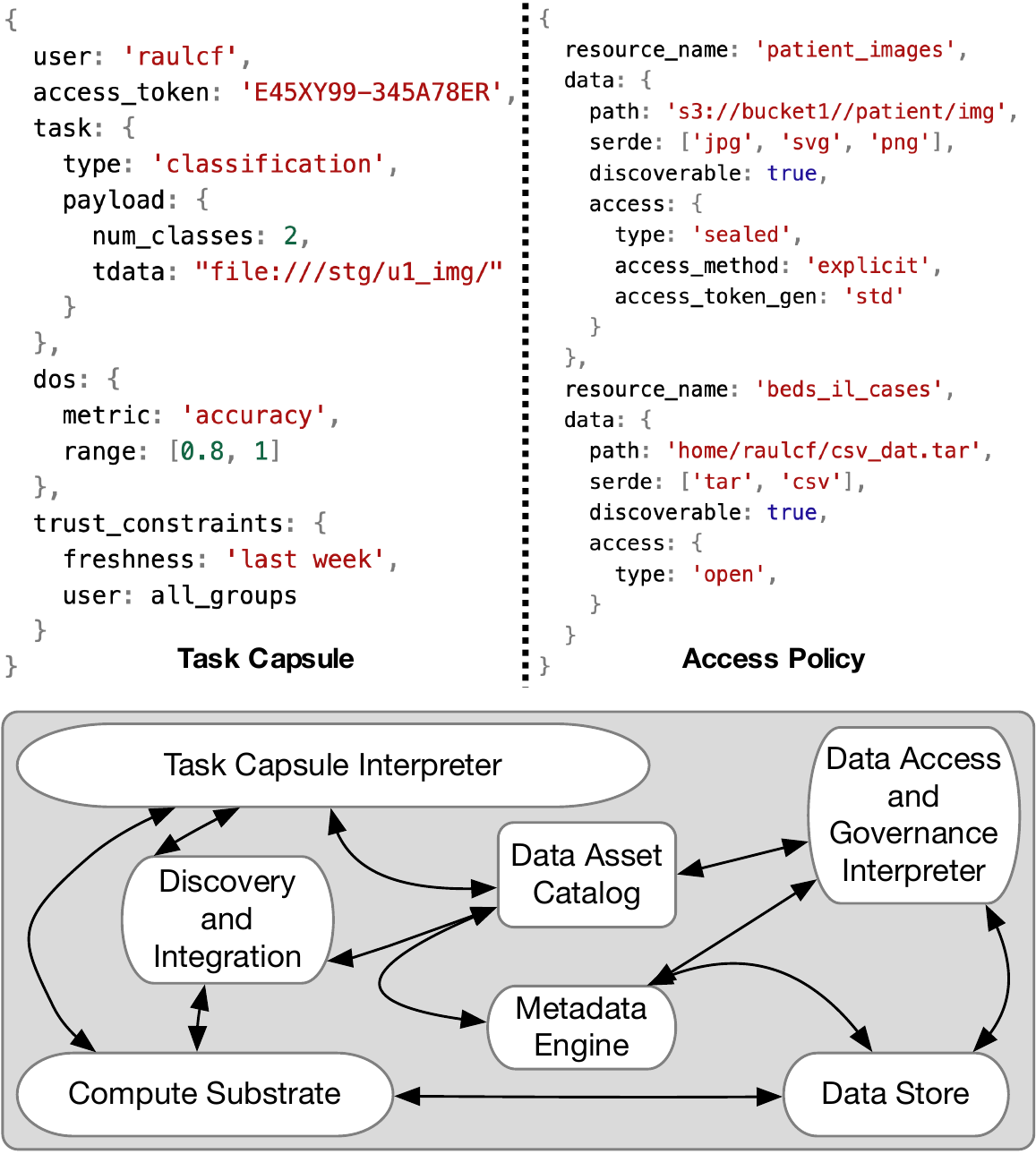} 

\vspace{-2ex}

\caption{Data Station architecture (bottom). Examples of Task capsule 
(top-left), and access policy (top-right).} 
\label{fig:architecture} 
\end{figure}

\subsubsection{Data User Perspective}
A Data Station does not deliver data to users; instead, 
users bring their computations to the
data. They do so by creating \textbf{data-unaware task capsules}. 
A \emph{capsule} encapsulates a
declaration of some computation to be performed, as well as criteria to verify
that the result is valid without looking at the data first.
A capsule is said to be \emph{data-unaware} because 
users have no access to any data when they
create a capsule.
As illustrated in the top left of  
Figure~\ref{fig:architecture}, a task capsule definition has three components:

%\begin{figure}[h] 
%\centering
%\includegraphics[width=\columnwidth]{img/taskcapsule.pdf} 
%\caption{Example Task capsule definition.}
%\label{fig:taskcapsule} 
%\end{figure}

%\begin{verbatim}
%{
%  user: 'raulcf',
%  access_token: 'E45XY99-345A78ER',
%  task: {
%  	type: 'classification',
%    payload: {
%      num_classes: 2,
%      tdata: "file:///stage/u11/hinge_img/"
%    }
%  },
%  dos: {
%  	metric: 'accuracy',
%    range: [0.8, 1]
%	},
%  trust_constraints: {
%    freshness: 'last week',
%    user: all_groups
%  }
%}
%\end{verbatim}

\begin{myitemize} 

\item \textbf{Task specification.} A task specification consists of a \emph{task
type} that selects a computational task from a extensible finite set, \eg
classification, and a \emph{task payload} that includes type-dependent
information. The example task capsule indicates there are two classes and 
specifies a path to test data.

%\item \textbf{Task specification.} A data specification consists of a \emph{task
%type} and a \emph{task payload}. The task type indicates the computational task,
%\eg, \emph{classification}. Data Stations support a finite set of task types
%that can be extended. The task payload depends on the task type.  The example
%task capsule indicates there are two possible classes and includes a path to the
%test data.  

\item \textbf{Degree of satisfaction (DOS)}. This metric depends on the task
type and is used to determine what results are valid to users, \eg demanding a
ML model accuracy to be $\ge 0.8$, as in the example.

\item \textbf{Trust constraints}. To trust the results, users want to know what
datasets contributed to the result and when, by whom, and how the dataset was created.
%by whom, and by which program. 
Lacking access to data, users cannot verify these criteria directly.
directly. Instead, they include these requirements (see the example) in the form
of constraints that are checked by the Station before delivering results.

\end{myitemize}

\mypar{Task Capsule Types} Data Stations can support other task types besides
classification, such as \emph{Query-by-Example}~\cite{qbe} interfaces for
analytical queries, ML tasks such as regression and anomaly
detection via autoML~\cite{automl}, and search.

\subsubsection{Data Contributor Perspective}
\label{subsubsec:dc}

When a data contributor uploads a dataset to the Station they include a
\emph{signature}---based on public key cryptography---that identifies
them as owning and being responsible for the dataset. By default, only a dataset's owner(s)
is granted access; the dataset remains otherwise invisible to all
other users. Any further access to the dataset, or to any dataset derived from
the dataset, must be mediated. We explain the protocol later in the section
and focus now on the policy.

To make accessible by others, owners declare an \textbf{access policy} (see example in
\F\ref{fig:architecture}, top-right) that includes a minimum of three
properties: \emph{discoverability}, \emph{access}, and \emph{derivation}.
\emph{Discoverability} indicates whether the discovery module can include the
dataset in responses to searches. \emph{Access} indicates whether the dataset is
closed to everyone (the default), open to everyone, or brokered; the latter case indicates
that explicit permission must be given before the dataset can be accessed.
Finally, \emph{derivation} (not shown in the example)
%\kyle{I dont see this in the figure?,maybe should state that its not shown?} 
indicates whether the dataset can be combined with
others or offered as-is. More fine-grained controls are also supported.
For example, an access policy may give access to a relational dataset only by tasks
of type \emph{analytical} (queries with joins, group by,
aggregations), only through a differential
privacy~\cite{differentialprivacy} filter, and with the number of
accesses constrained to control the
%\kyle{differential privacy?} 
\emph{privacy budget}, $\epsilon$. (The $\epsilon$
parameter controls
the privacy in differential privacy.) Such
access policies permit contributors to bound the access and usage of
datasets without engaging in complex data sharing agreements.

\mypar{Bulk uploads} Data contributors can upload entire data systems to the
Station at once---such as when unlocking silos---and include a default access
policy that applies to every dataset.
%
%\mypar{Updating datasets} 
The Station provides tools and APIs to update datasets
previously submitted. 
%All previous policies remain valid for the newly updated
%data.

\mypar{Encrypted datasets} Data contributors can upload encrypted datasets to
comply with certain regulations, as long as those datasets are accompanied by non-encrypted
metadata---that is, the metadata that would normally be extracted by the metadata engine and from
humans via incentives.

%After overviewing the data user and contributor interface with the Station, we
%now present some of the critical components at the center of Data Stations.

\subsection{Station Workflow}\label{subsec:1}

The Data Station component diagram is shown at the bottom of
\F\ref{fig:architecture}. Upon receiving a task capsule, the Data Station uses
(\textbf{Step 1}) a discovery platform to identify datasets that are potentially
relevant to the task, and then (\textbf{Step 2}) an integration platform to
combine datasets so they are valid inputs to the capsule. It then allocates
compute (\textbf{Step 3}) to evaluate the task on those datasets, checking
whether any of them satisfy the DOS metric, \eg the accuracy for an ML model.
When a solution is found, the Station (in \textbf{Step 4}) interacts with the
data user to mediate access to the results. Because solving this problem
automatically is not always possible, Stations incentivize humans to participate
when and where they are most needed. Stations use market mechanisms, which we
explain in Section~\ref{sec:market} to achieve this goal.

\noindent\textbf{Step 1: Data Discovery.} The goal of data discovery is to
identify datasets that are \emph{relevant} to a task capsule among thousands of
diverse heterogeneous datasets. A dataset is relevant if it helps solve the task
and it satisfies the capsule's constraints. The discovery module is based on
Aurum~\cite{aurum}, and it uses a \emph{data asset catalog} and \emph{discovery
indexes}. The catalog maintains the lifecycle of each dataset in the Station in
the form of \emph{profiles}, which are descriptions of the data, such as
statistical distribution of values, sketches, but also temporal information and
others. Profiles are automatically computed by the Metadata Engine when datasets
are submitted to the Station, or elicited from humans using incentives.
Discovery indexes are built from the catalog to ease dataset search and include
similarity indexes to find complementary data, full-text search indexes to match
keywords, linkage graphs to identify potential join paths---in the case of
relational data, and many others. Any information in capsules useful to search
data is used to query the discovery indexes and catalog. For example, the trust
constraints are verified against the catalog to quickly prune potential results.
Then, depending on the task type, test data is used to find training data
(similar data) for building models, or, when the task type is QBE, attributes
and data samples are used to steer the search. 

\noindent\textbf{Step 2: Data Integration and Blending.} The goal 
is to transform a list of input datasets (the output of discovery) into a
desired output dataset (the input of a task capsule). Blending uses techniques
from program synthesis~\cite{programsynthesis}, ML, and others, to
identify what preparation and integration steps are necessary to derive the
output from the input. Bounding the number of task types supported is needed to
design blending engines tailored to analytical queries, ML tasks
that require training data, etc. Some of the techniques used by this module
include identifying mapping and transformation functions to join attributes
(\ie, in the case of relational data) as well as normalization and
standardization tasks, such as value interpolation to join on different time and
space granularities. 

In the next steps, Stations search for a pair of (task capsule, blending output
dataset) that satisifies the capsule's DOS (\textbf{Step 3}). Access to task
results and derived data products is mediated in \textbf{Step 4}. We explain
both steps in Sections~\ref{subsec:gov} and \ref{subsec:unbounded}. We start by
describing important components of the Station.

%In the next steps, Stations evaluate all pairs of (task capsule, blending output
%dataset) searching for one that satisfies the capsule's DOS (\textbf{Step 3}).
%Access to task results and derived data products---such as when requesting a ML
%inference---is mediated in \textbf{Step 4}. We explain both steps in
%Sections~\ref{subsec:gov} and \ref{subsec:unbounded}. Before that, we further
%describe important components of the Station.

\subsection{Data Asset Catalog}
\label{subsec:catalog}

The data asset catalog maintains the lifecycle and \emph{metadata}---how data
came to be, how it changed, how it's been used---of each dataset and derived
dataset hosted in the Station. The catalog serves the discovery and integration
engine, by data users to describe capsule's \emph{trust constraints}, and by
contributors and stewards to implement access and governance policies.

%At the core of the Data Station is a catalog that keeps track of each dataset
%and derived data product's lifecycle, from how it came to be, to how it has
%changed, its purpose, where it lives and comes from, and machine as well as
%human-provided descriptions. The catalog is used by the Station's discovery and
%blending engines to identify candidate datasets for task capsules.  It is used
%by data users to describe \emph{trust constraints} as part of task capsules,
%which is critical since they cannot see the datasets otherwise.  Finally, the
%catalog also keeps and manages access and governance policies that data
%contributors submit.

To be interpretable by data users, data contributors, and the Station itself, the
catalog implements a common mental model consisting of \emph{profiles} and
\emph{relationships}. Profiles include: \emph{what-profile} to describe an actual
dataset, \emph{how-profile} to indicate what program produced the current
dataset version, \emph{who-profile} to indicate who produced and who uses the
dataset, \emph{where-profile} to indicate how the dataset can be accessed,
\emph{why-profile} to explain the purpose of the dataset, and
\emph{when-profile} to explain when the dataset was modified and when it is
valid. Relationships are built out of profiles: for example, provenance is built from
who- and how-profiles, and syntactic relationships such as join and similarity
graphs are built from what-profiles. Both profiles and relationships are used by
the discovery and integration modules.

%Because it needs to be interpreted by data users, contributors, and the Station
%itself, the catalog must implement a mental model comprehensible to all parties.
%It maintains different dataset \emph{profiles}: \emph{what-profile} to describe
%the actual dataset, \emph{how-profile} to indicate what program produced the
%current dataset version, \emph{who-profile} to indicate who produced and who
%uses the dataset, \emph{where-profile} to indicate how the dataset can be
%accessed, \emph{why-profile} to explain the purpose of the dataset, and
%\emph{when-profile} to explain when the dataset was modified and when is it
%valid. The catalog also maintains \emph{relationships} between datasets, which
%are computed based on each profile. For example, \emph{provenance} is computed
%based on how- and who-profiles, providing a trace of what programs and people
%contributed to the creation of datasets. They include syntactic relationships
%(based on what-profiles) which are used, among others, by the discovery module
%to build discovery indexes such as similarity between data assets, and others.
%All these relationships are important to identify relevant datasets, as well as
%for data users to build trust on results; \ie certain users may not want data
%derived from certain datasets.

The catalog's logical schema design strives for a balance between
\emph{structuredness}, which facilitates querying, and \emph{flexibility}, which facilitates
including new data. At its core, it reflects the mental model introduced
above, which allows different parties to understand and query it effectively.
To increase flexibility, it supports semi-structured data, such as JSON, to
reflect the idiosyncrasies of different data formats: \eg describing an image is
different than describing a relation or a ML model.

\mypar{Populating the Catalog} The Data Station triggers the execution of a
metadata engine whenever a new dataset is received, an existing dataset is updated,
or the Station produces a derived data product from existing ones. The metadata
engine analyzes the dataset and extracts as many \emph{profiles} as it can
automatically. This is done via the orchestration of analyzers that specialize
in different profiles, but also by eliciting this information from data users
and contributors directly when it cannot be accessed differently (see
Section~\ref{sec:market}. As a consequence, the full lifecyle of residing
datasets and derived data products is known because all operations on that
dataset happen within the realm of the Station. 

\mypar{Catalog Service} A catalog service facilitates loading and querying, and
stores and enforces access and governance policies. The service maintains
schemas of the semi-structured data as well as indexes that permit the discovery
and blending engines to find the information they need, and data users to
specify trust constraints to guarantee the origin and nature of the results they
request.

\subsection{Scalable Governance and Access} 
\label{subsec:gov}

We consider two types of data policies: \textit{data governance policies}
control what derived data products are produced in the Station while
\textit{data access policies} indicate who can access what data. 

\mypar{Data governance policies} These specifications limit and control 
the use of datasets and data tasks. For example, one may want to prohibit
production of derived data from datasets that contain personally identifiable information (PII). If PII is
defined specifically enough, for example, by providing a table with 116
attributes that correspond to PII data, then the \emph{metadata engine} can tag
datasets that contain such information by matching the definition with the
existing data---represented in what-profiles.

All governance policies are registered with the Catalog service, which is responsible 
for their enforcement. Only datasets and derived data products that pass
the constraints are returned as a result. Data governance policies not only
apply to data. They also govern what task capsule implementations are permitted,
for example what kind of ML models are used. This permits prohibiting the use of
certain models that are not sufficiently interpretable, or that are susceptible
to biases otherwise.

Finally, because all data lifecycles are known to the Station via the catalog,
it is possible to specify and enforce governance policies that apply actively to existing
data, \eg removing datasets and derived data products subject to the `right to
be forgotten'~\cite{rtbf}. This capability facilitates complying with regulations such as
GDPR and CCPA within the realm of the Station.

\mypar{Data access protocols} Sealing all data and derived data
products mitigates many of the problems associated with sharing sensitive
information, but it requires implementation of a solution that will allow
autorized users to access the results of their computations. The Data Station
architecture is amenable to capability-based mechanisms~\cite{capabilitybased1,
capabilitybased2} that give access to results as long as the computation
includes an adequate \emph{access token}. Access tokens can be requested from
the platform or directly from the data contributors. The choice of mechanism can be
left up to the preference of the data contributor and based on different
\emph{access policies}. 
%For example, if a data contributor decides that a dataset is freely available, 
%the Station will provide an access token to data users. Alternatively, if the
%contributor prefers to individually vet access, every request must be approved
%by the contributor to obtain an access token.

Access tokens can encapsulate richer information than merely a boolean value that
indicates if access is granted. For example, they can grant one-time access or alternatively provide
an expiry date---which can be infinite when access is granted permanently. Once
in possession of a token, users can seamlessly use the Station to work on those data; they
need only request new access token when accessing new and protected
datasets. The burden of creating and managing tokens is on the Station, which
understands what datasets have contributed to the results being requested and
hence can orchestrate the actions needed to grant, manage, and revoke access as required.

\subsection{Dealing with Unbounded Computation}
\label{subsec:unbounded}

In a traditional data processing architecture, such as a database, a task consists
of a query that expresses the computation to perform and the data to be read.
Given these two pieces of information, it is often possible to estimate the computational resources
needed to obtain the results. This is not true in Data Stations, where the goal
is to identify datasets for which the defined task achieves the desired DOS
metric. Thus, the amount of computation that may be required to perform a user
request may be ivirtually unbounded, as it may be necessary to
check all dataset combinations.

Data Stations rely on two main mechanisms to tackle this challenge. First,
the discovery and integration platform enable srapid pruning of the space of
compatible datasets. Second, the Data Station is 
implemented on a modern execution platform (e.g., cloud) with the
scalability and performance required to solve manyf tasks. 
A result cache maintains the relationship
between executed tasks and datasets accessed, with the goal of
informing and guiding the matching of tasks
with data in the future, which is done, in turn, via a scheduler in a
speculative manner.  

Finally, Data Stations are only logically centralized. Physically, computation
may take place across multiple machines that may be dedicated to specific tasks,
such as serving ML models that are otherwise not accessible to users. Each
compute node is stateless and accesses data from a disaggregated data store;
dissaggregation of compute and storage facilitates scalability.

\subsection{How Stations Address Requirements}
\label{subsec:summary}

The challenges (\textbf{C1-C4}) are addressed by the task capsules, the discovery
and blending module, and the unbounded computation engine, as well as the catalog and
its ability to manage and enforce access and governance policies.

By capturing and maintaining the lifecycle of each dataset and derived data
product, the data asset catalog along with the discovery and blending engines
satisfy \textbf{REQ1}, \textbf{REQ2}, and \textbf{REQ3}.  The Station
architecture, by sealing data and mediating access through fine-grained
policies, helps with \textbf{REQ4}.

Nevertheless, Stations cannot solve all problems automatically. Semantic ambiguity in
discovery and integration tasks~\cite{discoveryambiguity},
for example, require humans in the loop. A metadata engine keeps the catalog up
to date, but certain profiles are impossible to create automatically and need
human input, \eg a why-profile that describes the reason for the existence of a
dataset. Finally, even if data in the Station are technically secured, humans may
have other disincentives to share the data, such as fear of leaking proprietary
information. This section has dealt with the technical problem, we
discuss in the next section how Data Stations host data markets to help manage
incentives to tackle the human factor.

\section{Market Forces and Incentives}
\label{sec:market}

During the course of processing a task capsule, the Station may run into
situations where it requires human input to make progress: for example,
when it needs to join two tables on an attribute \textsf{address}, but lacks 
the information to choose between two
alternatives, \textsf{work address} and \textsf{home address}.  Stations may
also block because they cannot determine if an action is safe. For example, a
capsule wants to train an ML model and the Station has identified a candidate
training dataset with sample data, but it cannot tell how the sample was
generated and the corresponding \emph{why-profile} is incomplete. 

Data Stations introduce incentive mechanisms to motivate data contributors and
data users to treat data as a valuable asset and help solve data problems when
the technical solution is insufficient. Stations coordinate human effort to
curate, document, and prepare data when ambiguity in task capsules, catalog,
schemas, or other data descriptions prevents progress:
\eg by incentivizing the creator of the ML training dataset in the example above to
explain how the data sample was collected. This coordination of human and
machine is achieved via data market designs and relies on two mechanisms:
task generation and incentives.

\emph{Task generation.} When Stations block on a task they create human-readable
task descriptions. These task descriptions must incorporate sufficient context so
humans can effectively solve the problem. \emph{Incentives.} When assigning a
task to humans, Stations must indicate the incentives humans will receive in
exchange for solving the problem. Incentives can be currencies of different
kinds, such as money, time tokens, or others. For example, the Station may
create a task that request filling the \emph{why-profile} of the example above
in exchange for 30 minutes of leisure time.

\mypar{Balancing incentives and utility} Each participant seeks to maximize its
utility model. For example, the participant of the example above will take on
the task of filling the \emph{why-profile} if they perceive the gain, 30
minutes, to be more valuable to them than the effort needed to complete the
profile. It is safe to assume that data users maximize their utility when
Stations answer their task capsules fast. Stations must account for the utility
gained by data buyers, the one gained by data contributors, and strike an
equilibrium that maximizes the utility of the market as a whole.

\mypar{On designing data markets} Internal data markets such as that
described above differ in their characteristics from those needed for a Station
that serves a consortium of entities, such as a group of hospitals. An individual's
motivations and the Station's goals are different, as are the levels of trust
among entities. These differing qualities, in turn, call for different market
designs. In order to design markets for scenarios where participants behave
strategically to maximize their utility, we design truthful
mechanisms~\cite{mechanismdesign} to align participants' incentives with
Station goals.
%\smallskip
Incentivizing those more familiar with the data allows Data Stations to solve
task capsules while keeping all data sealed by default.

\section{Discussion and Related Work}
\label{sec:relatedwork}

As analysts ask more varied questions, the schema-first
approach of warehouses becomes a limiting factor. Data lakes~\cite{lake1, lake2,
kudu, hudi, delta} are a partial answer to this problem. Lakes store the data
first and push the burden of interpreting schemas to end users. By doing so,
data lakes worsen the discovery and integration problem. Data Stations depart
from traditional data architectures in different ways. They make the data asset
catalog a crucial component of their architecture and make sure it is up to date
at all times. The catalog, in turn, powers the discovery and integration engine
which solves these problems without the need for agreeing on a schema a priori,
such as in federated integration systems~\cite{garlic1, garlic2}. Although
automatically solving discovery and integration problems is difficult, Stations
take advantage of its logical centralization of data and compute to implement
market structures that incentivize humans to get in the loop and solve the
hardest problems at its root, similar to what Anylog~\cite{anylog} does for
distributed \emph{IoT} scenarios. 

%In prior work we
%explored the legal, technical, and operational difficulties as well as the
%state-of-the-art approaches for managing sensitive
%data~\cite{foster2018research}. We concluded that there is a need for a hosted
%data enclave, similar to the Data Station, for providing safe interaction for
%analysts, data, and software, and as a means of automating and thus
%professionalizing data stewardship processes.

\mypar{An opportunity for data management} Data management problems remain as
hard and relevant as ever, with human and technical factors that are uniquely
shaped by the challenges of our time: ever increasing volumes of data that are
hoarded by a few, and that are difficult to share for technical and legal reasons alike.
New applications such as ML and statistical methods demand
new query interfaces, and introduce new ethical problems due to their rapidly
increasing impact in our lives.  Increasing awareness of the role of
data in society is leading to new regulations and laws. We believe it is
time to rethink data architectures to tackle these modern challenges. Data
Stations are a step towards this goal.

\bibliographystyle{abbrv}
\balance
\bibliography{main}

\end{document}